**In-Silico Analysis of Curve Fitting in Angiographic Parametric Imaging in Intracranial Aneurysms**
Parmita Mondal[1,2], Allison Shields[1,2], Mohammad Mahdi Shiraz Bhurwani[3], Kyle A Williams[1,2], Ciprian N Ionita[3]
[1]Department of Biomedical Engineering, University at Buffalo, Buffalo, NY 14260
[2]Canon Stroke and Vascular Research Center, Buffalo, NY 14203
[3]QAS.AI Inc, Buffalo, NY 14203

**Abstract**

**Purpose:**
In Angiographic Parametric Imaging (API), accurate estimation of parameters from Time Density Curves (TDC) is crucial. However, these estimations are often marred by errors arising from factors such as patient motion, procedural preferences, image noise, and injection variability. While fitting methods like gamma-variate fitting offer a solution to recover incomplete or corrupted TDC data, they might also introduce unforeseen biases. This study investigates the trade-offs and benefits of employing gamma-variate fitting on virtual angiograms to enhance the precision of API biomarkers.

**Materials and Methods:**
Utilizing Computational Fluid Dynamics (CFD) in patient specific 3D geometries, we generated a series of high-definition virtual angiograms at distinct inlet velocities: 0.25m/s, 0.35m/s, and 0.45m/s. These velocities were investigated across injection durations ranging from 0.5s to 2.0s. From these angiograms, TDCs for aneurysms and their corresponding inlets were constructed. To emulate typical clinical challenges, we introduced noise, simulated patient motion, and generated temporally incomplete data sets. These modified TDCs underwent gamma-variate fitting. We quantified both the original and fitted TDC curves using standard angiography metrics such as Cross-Correlation (Cor), Time to Peak (TTP), Mean Transit Time (MTT), Peak Height (PH), Area Under the Curve (AUC), and Maximum Gradient (Max-Gr) for a comprehensive comparison.

**Results:**
TDCs enhanced by gamma-variate fitting exhibited a robust correlation with vascular flow dynamics. Our results affirm that gamma-variate fitting can adeptly restore TDCs from fragmentary sequences, elevating the precision of derived API parameters.

**Conclusions:**
Incorporating gamma-variate fitting into TDCs analysis augments the precision and robustness of API parameters, bolstering the credibility of neurovascular diagnostic procedures.

**BACKGROUND**

Endovascular intervention for intracranial aneurysms (IAs) is at the forefront of neuro-interventional advancements. IAs, dilations within brain blood vessels, present a significant risk. If ruptured, they can cause subarachnoid hemorrhage, leading to severe outcomes, making the need for precision in therapeutic tools essential. While the advent of modern endovascular devices and methodologies has enhanced IA treatment outcomes, challenges persist. There isn't a universally accepted metric to anticipate treatment results intraoperatively. Consequently, clinicians heavily rely on their expertise and immediate angiographic interpretations. Angiographic Parametric Imaging (API) has emerged as a promising tool in this regard. Beyond merely visualizing the vascular nuances, API translates these images into actionable, dynamic data for better interventions. This technique, aimed at refining angiographic data, seeks to provide clinicians with dependable real-time data. It focuses on ensuring that decisions are based on the most accurate insights. Preliminary investigations into API based on Time Dilution Curves (TDC) parametrization using metrics such as Cross-Correlation (Cor), Time to Peak (TTP), Mean Transit Time (MTT), Peak Height (PH), Area Under the Curve (AUC), and Maximum Gradient (Max-Gr) Figure 1, have shown real potential. Furthermore, by using the TDC, a quantitative description of contrast for each voxel is generated.

However, the crucial data from angiograms is frequently compromised. Factors such as patient movement, procedure-specific variations, image noise, and contrast injection inconsistencies can distort the clarity of angiograms. These aren't just trivial issues; they can seriously hamper the angiogram's interpretability, potentially leading to clinical misjudgments. In an environment devoid of reliable predictive metrics, the importance of restoring corrupted angiographic data becomes vital for optimizing IA treatment outcomes. Gamma-variate fitting has emerged as a potential solution, equation 1, which is computationally inexpensive, and relatively transparent [1].

$$C_{GI}(t; \alpha, \beta) = t^{\alpha-1} e^{-\beta t} \beta^\alpha \qquad (1)$$

where t represents time and (α) and scale (β) are fitting parameters.

In practice, TDCs chart the variation in contrast agent concentration within a Region of Interest (ROI) during an imaging session. ROIs are manually delineated for each aneurysm and its associated inlet in angiograms. There are instances where an interventionalist might prematurely halt x-ray imaging, missing the TDC's "tail". On occasions, especially when patients are awake during the procedure, they might move. Additionally, certain orientations of the C-arm could degrade the contrast's signal-to-noise ratio. It's understood that gamma-variate fitting of TDCs essentially serves as a smoothing filter for angiographic data [2, 3].

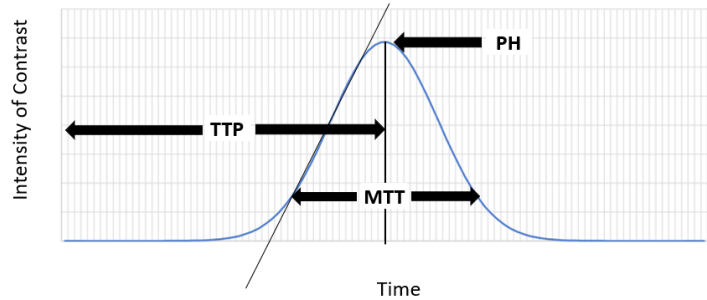

**Figure 1.** Plot of a Time density Curve(TDC) showing Time to Peak(TTP), Mean Transit Time(MTT) and Peak Height(PH). The blue line shows the TDC while the black arrows exemplify the TTP, MTT and PH.

Given these scenarios, it's reasonable to question the accuracy of data recovered through gamma fitting [4]. It also raises concerns about potential biases introduced by this method and whether specific considerations should be incorporated.

This study delves into detail used of gamma-variate fitting, using virtual angiography to investigate the accuracy of the method and potential biases.

**Material and Methods**
*Data Collection*
Data collection and analysis was approved by our Institutional Review Board. The **Figure 2** shows chronologically the steps taken to complete this study. Virtual angiograms were generated in Computational Fluid Dynamics (CFD) for inlet velocities 0.25m/s, 0.35m/s and 0.45m/s with each having an injection duration of 0.5s, 1.0s, 1.5s and 2.0s [5]. These virtual angiograms give the structural assessment of the vasculature and helps in visualizing contrast propagation in the blood vessels and its surrounding environment. We generated TDCs for the aneurysms and their respective inlets in the vasculature, from these virtual angiograms using gamma-variate fitting. We then extracted API parameters, namely Cor, TTP, MTT, PH, AUC, and Max-Gr from the TDCs of the aneurysm and their inlet. API is a semiquantitative angiographic tool that generates a set of vascular maps, where each pixel is colored according to the intensity of a certain flow parameter [6]. To reduce hand injection variability, we normalized each parameter to their corresponding inlet values to yield NI-Cor, NI-TTP, NI-MTT, NI-PH, NI-AUC, and NI-Max-Gr, respectively, where NI is normalized to inlet. Normalized parameters were analyzed in Statistical Package for Social Sciences (SPSS) using standard ROC analysis.

In this study, we consider three different cases to study the effect of gamma-variate fitting on virtual angiograms. In the first case, we consider 80% of the image sequences in the virtual angiogram and then fit it with gamma-variate, to extract the API parameters from its TDC. In the second case, we add Poisson noise to the virtual angiogram and then fit it with gamma-variate, to extract the API parameters from its TDC. In the last case, we shift the pixels in the virtual angiogram slightly, to account for patient motion. We fit the TDC with gamma-variate and extract the API parameters from it. This helps in studying the API parameters of the original TDC vs gamma-variate fitted TDC.

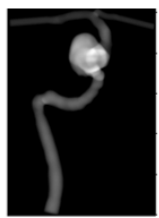 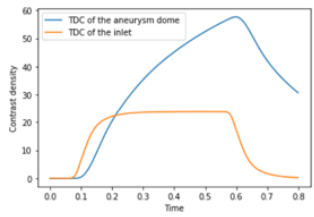 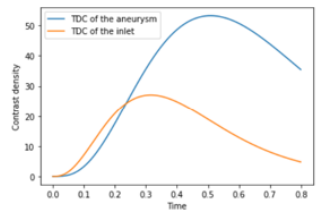

| API Parameters | Values |
|---|---|
| NI-Cor | 1.9 |
| NI-MTT | 1.54 |
| NI-TTP | 1.53 |
| NI-PH | 1.97 |
| NI-AUC | 2.90 |
| NI-MaxGr | 1.257 |

The phantom above is generated from Computational Fluid Dynamics (CFD). Here we have taken 80% of the image sequences into consideration.

The Time Density Curve(TDC) of the aneurysm dome and the inlet is shown above before gamma-variate fitting.

The Time Density Curve(TDC) of the aneurysm dome and the inlet is shown above after gamma-variate fitting

API parameters extracted from the Time Density Curve(TDC) after gamma-variate fitting is shown above

A

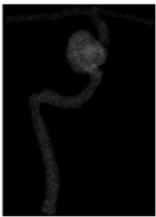 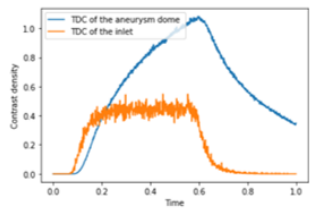 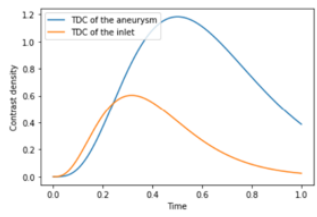

| API Parameters | Values |
|---|---|
| NI-Cor | 1.75 |
| NI-MTT | 1.522 |
| NI-TTP | 1.514 |
| NI-PH | 1.94 |
| NI-AUC | 2.86 |
| NI-MaxGr | 1.27 |

The phantom above is generated from Computational Fluid Dynamics (CFD). Here we have added Poisson noise to the image sequences.

The Time Density Curve(TDC) of the aneurysm dome and the inlet is shown above before gamma-variate fitting.

The Time Density Curve(TDC) of the aneurysm dome and the inlet is shown above after gamma-variate fitting

API parameters extracted from the Time Density Curve(TDC) after gamma-variate fitting is shown above

B

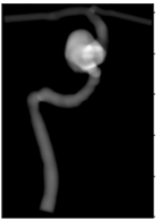 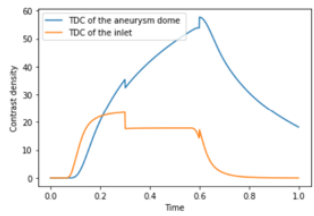 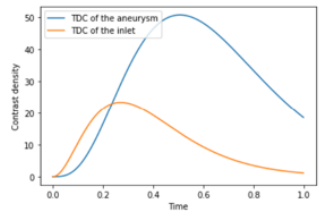

| API Parameters | Values |
|---|---|
| NI-Cor | 3.02 |
| NI-MTT | 1.65 |
| NI-TTP | 1.67 |
| NI-PH | 2.17 |
| NI-AUC | 3.19 |
| NI-MaxGr | 1.22 |

The phantom above is generated from Computational Fluid Dynamics (CFD). Here, we have shifted the pixels slightly to account for patient motion.

The Time Density Curve(TDC) of the aneurysm dome and the inlet is shown above before gamma-variate fitting.

The Time Density Curve(TDC) of the aneurysm dome and the inlet is shown above after gamma-variate fitting

API parameters extracted from the Time Density Curve(TDC) after gamma-variate fitting is shown above

C

**Figure 2.** Workflow showing chronologically the steps taken to complete the study. A. In this case, we take into consideration 80% of the image sequences in the virtual angiogram and then gamma-variate fit the Time density Curve(TDC) to extract the Angiographic Parametric imaging(API) parameters. B. In this case, we add Poisson noise to the virtual angiogram and then gamma-variate fit the Time density Curve(TDC) to extract the Angiographic Parametric imaging(API) parameters. C. In this case, we shift the pixels in the virtual angiogram to account for patient motion and then gamma-variate fit the Time density Curve(TDC) to extract the Angiographic Parametric imaging(API) parameters.

## Results

API data distribution extracted from gamma-variate fitted and non-gamma variate fitted TDCs are demonstrated using a box and Whisker plot in **Figure 3**.

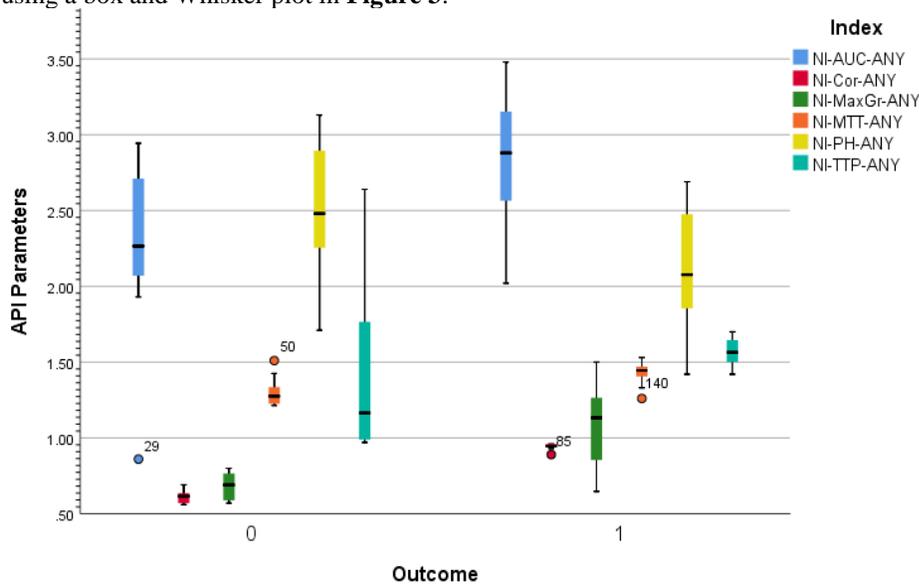

**Figure 3.** Whisker plots for API parameters: Cor, TTP, MTT, PH, AUC, and Max-Gr for gamma-fitted [1] and non-gamma fitted [0] TDCs. (ANY indicate aneurysm dome and NI is normalized to inlet

## Conclusion
This is an attempt to generate TDCs using gamma-variate fitting from virtual angiograms, for various inlet velocities and injection durations. This method enhances the accuracy and precision of API parameters.

## Acknowledgments
This work is supported by NSF STTR Award # 2111865